\documentclass[11pt,a4paper]{article}
\usepackage{a4wide}
\usepackage{mathpazo}
\usepackage{amsmath}
\usepackage{graphicx}
\usepackage[usenames,dvipsnames]{xcolor}
\usepackage[colorlinks=true,pagebackref=false,urlcolor=red,citecolor=blue,linkcolor=Emerald]{hyperref}
\usepackage[compress]{cite}
\usepackage{subfigure}
\usepackage{soul}
\usepackage{url}
\usepackage{tikz}
\usetikzlibrary{shapes.geometric, arrows}
\newlength{\fscale}
\newlength{\nscale}
\tikzstyle{process}=[rectangle, minimum width=3\fscale, minimum height=\fscale, text centered, draw=black, fill=yellow!30]
\tikzstyle{pout}=[process, fill=orange!30]
\tikzstyle{img}=[inner sep=0pt]

\tikzstyle{arrow}=[thick,->,>=stealth]

\newcommand{\G}{\mathcal{G}}

\newcommand{\X}{\mathbf{X}}

\newcommand{\eq}[1]{\begin{equation*}#1\end{equation*}}
\newcommand{\eqa}[1]{\begin{eqnarray}#1\end{eqnarray}}
\newcommand{\mat}[1]{\mathbf{#1}}

\newcommand{\cm}[1]{}

\newcommand{\cor}[1]{\mathrm{cor}(#1)}

\title{Detection of regulator genes and eQTLs in gene networks}

\author{Lingfei Wang$^{1}$ and Tom Michoel$^{1,\ast}$}

\date{}

\begin{document}

\maketitle

$^1$Division of Genetics and Genomics, The Roslin Institute, The
University of Edinburgh, Midlothian EH25 9RG, Scotland, United Kingdom

\medskip

$^\ast$Corresponding author, E-mail: tom.michoel@roslin.ed.ac.uk

\medskip

\begin{abstract}
  Genetic differences between individuals associated to quantitative
  phenotypic traits, including disease states, are usually found in
  non-coding genomic regions. These genetic variants are often also
  associated to differences in expression levels of nearby genes (they
  are ``expression quantitative trait loci'' or eQTLs, for short) and
  presumably play a gene regulatory role, affecting the status of
  molecular networks of interacting genes, proteins and
  metabolites. Computational systems biology approaches to reconstruct
  causal gene networks from large-scale omics data have therefore
  become essential to understand the structure of networks controlled
  by eQTLs together with other regulatory genes, as well as to generate
  detailed hypotheses about the molecular mechanisms that lead from
  genotype to phenotype. Here we review the main analytical methods
  and softwares to identify eQTLs and their associated genes, to
  reconstruct co-expression networks and modules, to reconstruct
  causal Bayesian gene and module networks, and to validate predicted
  networks \textit{in silico}.
\end{abstract}

\tableofcontents

\section{Introduction}
\begin{figure}[p]
\includegraphics{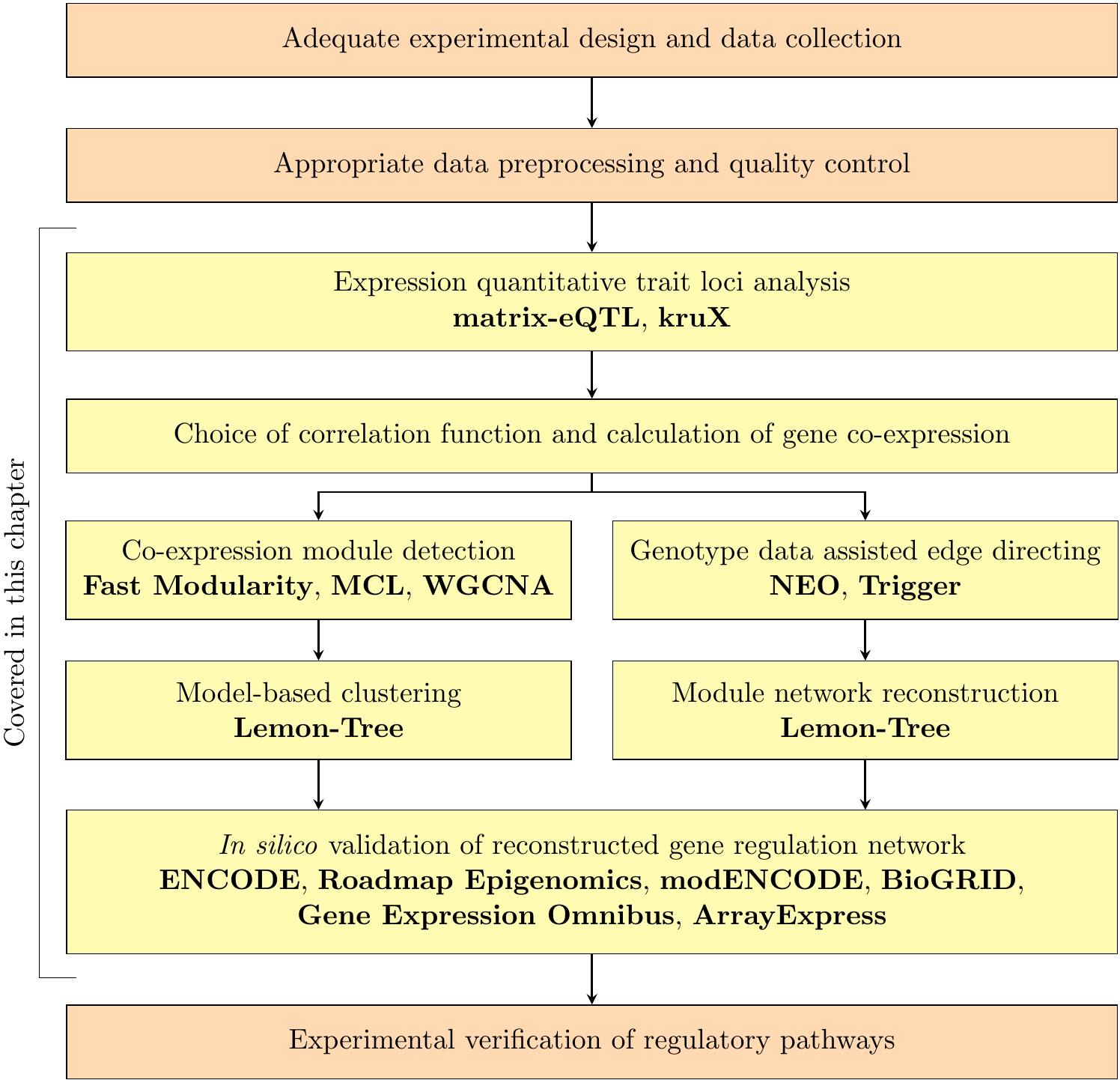}
\caption{A flow chart for a typical systems genetics study and the corresponding softwares. Steps in light yellow are covered in this chapter.\label{fig-flow}}\end{figure}

Genetic differences between individuals are responsible for variation
in the observable phenotypes.  This principle underpins genome-wide
association studies (GWAS), which map the genetic architecture of
complex traits by measuring genetic variation at single-nucleotide
polymorphisms (SNPs) on a genome-wide scale across many individuals
\cite{mackay2009genetics}. GWAS have resulted in major improvements in
plant and animal breeding \cite{goddard2009mapping} and in numerous
insights into the genetic basis of complex diseases in human
\cite{manolio2013bringing}.  However, quantitative trait loci (QTLs)
with large effects are uncommon and a molecular explanation for their
trait association rarely exists \cite{mackay2009genetics}. The vast
majority of QTLs indeed lie in non-coding genomic regions and
presumably play a gene regulatory role \cite{hindorff2009potential,
  schaub2012linking}.  Consequently, numerous studies have identified
\textit{cis}- and \textit{trans}-acting DNA variants that influence
gene expression levels (i.e., ``expression QTLs''; eQTLs) in model
organisms, plants, farm animals and human (reviewed in
\cite{rockman2006genetics, georges2007mapping, cookson2009mapping,
  cheung2009genetics, cubillos2012lessons}). Gene expression
programmes are of course highly tissue- and cell-type specific, and
the properties and complex relations of eQTL associations across
multiple tissues are only beginning to be mapped
\cite{dimas2009common, foroughi2015, greenawalt2011survey, ardlie2015genotype}.  At the
molecular level, a mounting body of evidence shows that
\textit{cis}-eQTLs primarily cause variation in transcription factor
(TF) binding to gene regulatory DNA elements, which then causes changes
in histone modifications, DNA methylation and mRNA expression of
nearby genes; \textit{trans}-eQTLs in turn can usually be attributed
to coding variants in regulatory genes or \textit{cis}-eQTLs of such
genes \cite{albert2015role}.

Taken together, these results motivate and justify a systems
biological view of quantitative genetics (``systems genetics''), where
it is hypothesized that genetic variation, together with environmental
perturbations, affects the status of molecular networks of interacting
genes, proteins and metabolites; these networks act within and across
different tissues and collectively control physiological phenotypes
\cite{williams2006expression, kadarmideen2006genetical,
  rockman2008reverse, schadt2009, schadt2012new,
  civelek2014systems,bjorkegren2015genome}. Studying the impact of
genetic variation on gene regulation networks is of crucial importance
in understanding the fundamental biological mechanisms by which genetic
variation causes variation in phenotypes \cite{chen2008}, and is
expected to lead to the discovery of novel disease biomarkers and drug
targets in human and veterinary medicine \cite{schadt2009b}.  Since
direct experimental mapping of genetic, protein--protein or
protein--DNA interactions is an immensely challenging task, further
exacerbated by the cell-type specific and dynamic nature of these
interactions \cite{walhout2006unraveling}, comprehensive,
experimentally verified molecular networks will not become available
for multi-cellular organisms in the foreseeable future. Statistical
and computational methods are therefore essential to reconstruct
trait-associated causal networks by integrating diverse omics data
\cite{rockman2008reverse, schadt2009, ritchie2015methods}.

A typical systems genetics study collects genotype and gene, protein
and/or metabolite expression data from a large number of individuals
segregating for one or more traits of interest. After raw data
processing and normalization, eQTLs are identified for each of the
expression data types, and a co-expression matrix is constructed. Causal Bayesian
gene networks, co-expression modules (i.e. clusters)  and/or causal
Bayesian module networks are then reconstructed. \textit{In silico}
validation of predicted networks and modules using independent data
confirms their overall validity, ideally followed by experimental
validation of the most promising findings in a relevant cell line or
model organism (Figure \ref{fig-flow}). Here we review the main analytic
principles behind each of the steps from eQTL identification to
\textit{in silico} network validation, and present a selection of most
commonly used methods and softwares for each step. Throughout this
chapter, we tacitly assume that all data has been quality controlled,
pre-processed and normalized to suit the assumptions of the analytic
methods presented here. For expression data, this usually means
working with log-transformed data where each gene expression profile
is centred around zero with standard deviation one. We also assume
that the data has been corrected for any confounding factors, either
by regressing out known covariates and/or by estimating hidden factors
\cite{stegle2012using}.

\section{Genetics of gene expression}
\label{sec:genet-gene-expr}

A first step towards identifying molecular networks affected by DNA
variants is to identify variants that underpin variations in eQTLs of
transcripts \cite{cookson2009mapping}, proteins \cite{foss2007} or
metabolites \cite{nicholson2011genome} across individuals. When
studying a single trait, as in GWAS, it is possible to consider
multiple statistical models to explicitly account for additive and/or
dominant genetic effects \cite{laird2011}. However, when the possible
effects of a million or more SNPs on tens of thousands of molecular
abundance traits need to be tested, as is common in modern genetics of
gene expression studies, the computational cost of testing SNP-trait
associations one-by-one becomes prohibitive. To address this problem,
new methods have been developed to calculate the test statistics for
the parametric linear regression and analysis of variance (ANOVA)
models \cite{shabalin2012matrix} and the non-parametric ANOVA model
(or Kruskal-Wallis test) \cite{qi2014} using fast matrix
multiplication algorithms, implemented in the softwares
\textbf{matrix-eQTL}
(\url{http://www.bios.unc.edu/research/genomic_software/Matrix_eQTL/})
\cite{shabalin2012matrix} and \textbf{kruX}
(\url{https://github.com/tmichoel/krux}) \cite{qi2014}.

In both softwares, genotype values of $s$ genetic markers and
expression levels of $k$ transcripts, proteins or metabolites in $n$
individuals are organized in an $s\times n$ genotype matrix $\mat G$ and
$k\times n$ expression data matrix $\mat X$. Genetic markers take values
$0,1,\dots,\ell$, where $\ell$ is the maximum number of alleles
($\ell=2$ for biallelic markers), while molecular traits take
continuous values. In the linear model, a linear relation is tested
between the expression level of gene $i$ and the genotype value (i.e.\ the number of reference alleles) of SNP $j$. The corresponding test
statistic is the Pearson correlation between the $i$th row of $\mat X$ and
the $j$th row of $\mat G$, for all values of $i$ and $j$. Standardising the
data matrices to zero mean and unit variance, such that for all $i$ and $j$,
\begin{align*}
  \sum_{l=1}^nX_{il} = \sum_{l=1}^nG_{jl} = 0 \quad\text{and}\quad
   \sum_{l=1}^n X_{il}^2 = \sum_{l=1}^n G_{jl}^2 = n ,     
\end{align*}
it follows that the correlation values can be computed as
\begin{align*}
  R_{ij} = \sum_{l=1}^nX_{il}G_{jl} = (\mat X\mat G^T)_{ij},
\end{align*}
where $\mat G^T$ denotes the transpose of $\mat G$. Hence, a single matrix
multiplication suffices to compute the test statistics for the linear
model for all pairs of traits and SNPs.

The ANOVA models test if expression levels in
different genotype groups originate from the same distribution.
Therefore, ANOVA models can account for both additive and
dominant effects of a genetic variant on expression levels. In the
parametric ANOVA model, suppose the test samples are divided into
$\ell+1$ groups by the SNP $j$. The mean expression level for gene $i$ 
in each group $m$ can be written as
\begin{align*}
  \overline{X_i^{(m,j)}} = \frac1{n^{(m,j)}} \sum_{\{l\colon G_{jl}=m\}} X_{il},
\end{align*}
where $n^{(m,j)}$ is the number of samples in genotype group $m$ for SNP
$j$.

Again assuming that the expression data is
standardised, the F-test statistic for testing gene $i$ against SNP
$j$ can be written as
\begin{align*}
  F_i^{(j)} = \frac{n-\ell-1}{\ell} \frac{SS_i^{(j)}}{n-SS_i^{(j)}},
\end{align*}
where $SS_i^{(j)}$ is the sum of squares between groups,
\begin{align*}
  SS_i^{(j)} = \sum_{m=0}^\ell n^{(m,j)}\overline{X_i^{(m,j)}}^2.
\end{align*}

Let us define the $n\times s$ indicator matrix $\mat I^{(m)}$ for
genotype group $m$, i.e. $\mat I_{lj}^{(m)} = 1$ if $G_{jl}=m$ and $0$
otherwise. Then
\begin{align*}
  \sum_{\{l\colon G_{jl}=m\}} X_{il} = \left(\mat X\mat I^{(m)}\right)_{ij}.
\end{align*}
Hence, for each pair of expression level $X_i$ and SNP $G_j$,
the sum of squares matrix $SS_i^{(j)}$ can be computed via
$\ell-1$ matrix multiplications \footnote{There are only $\ell-1$ matrix 
multiplications, because the data
standardization implies that $\mat X\mat I^{(0)}=1-\sum_{m=1}^{\ell-1}\mat X\mat I^{(m)}$.}.

In the non-parametric
ANOVA model, the expression data matrix is converted to a matrix $\mat T$
of data ranks, independently over each row. In the absence of
ties, the Kruskal-Wallis test statistic is given by
\begin{align*}
  S_{ij} = \frac{12}{n(n+1)} \sum_{m=0}^\ell n^{(m,j)}\,\overline{T_i^{(m,j)}}^2 - 3(n+1),
\end{align*}
where $\overline{T_i^{(m,j)}}$ is the average expression rank of gene $i$
in genotype group $m$ of SNP $j$, defined as
\begin{align*}
  \overline{T_i^{(m,j)}}&=\frac1{n^{(m,j)}} \sum_{\{l\colon G_{jl}=m\}}T_{il},
\end{align*}
which can be similarly obtained from the $\ell-1$ matrix multiplications.

There is as yet no consensus about which statistical model is most
appropriate for eQTL detection. Non-parametric methods were introduced
in the earliest eQTL studies \cite{brem2002,schadt2008} and have
remained popular, as they are robust against variations in the
underlying genetic model and trait distribution. More recently, the
linear model implemented in matrix-eQTL has been used in a number of
large-scale studies
\cite{lappalainen2013transcriptome,ardlie2015genotype}. A comparison
on a dataset of 102 human whole blood samples showed that the
parametric ANOVA method was highly sensitive to the presence of
outlying gene expression values and SNPs with singleton genotype
group. Linear models reported the highest number of eQTL
associations after empirical False Discovery Rate (FDR) correction, with an expected bias
towards additive linear associations. The Kruskal-Wallis test
was most robust against data outliers and heterogeneous genotype group
sizes and detected a higher proportion of non-linear associations, but
was more conservative for calling additive linear associations than
linear models \cite{qi2014}.

In summary, when large numbers of traits and markers have to be tested
for association, efficient matrix multiplication methods
can be employed to calculate all test statistics at once,
leading to a dramatic reduction in computation
time compared to calculating these statistics one-by-one for every
pair using traditional methods.  Matrix multiplication is a basic
mathematical operation which has been purposely studied and optimized
for tens of years \cite{golub1996}. Highly efficient packages, such as
\textbf{BLAS} (\url{http://www.netlib.org/blas/}) and \textbf{LAPACK}
(\url{http://www.netlib.org/lapack/}), are available for use on
generic CPUs, and are indeed employed in most mainstream scientific
computing softwares and programming languages, such as Matlab and
R. In recent years, Graphics Processor Unit (GPU)-accelerated
computing, such as CUDA, has revolutionised scientific calculations
that involve repetitive operations in parallel on bulky data, offering
even more speedup than the existing CPU-based packages. The first
applications of GPU computing in eQTL analysis have already appeared
(e.g. \cite{hemani2014detection}), and more can be expected in the
future.

Lastly, for pairs exceeding a pre-defined threshold on the test
statistic, a $p$-value can be computed from the corresponding test
distribution, and these $p$-values can then be further corrected for
multiple testing by common procedures
\cite{shabalin2012matrix,qi2014}.

\section{Co-expression networks and modules}

\subsection{Co-expression gene networks\label{sec-coex}}
The Pearson correlation is the simplest and computationally
most efficient similarity measure for gene expression profiles.
For genes $i$ and $j$, their Pearson correlation can be written as
\begin{equation}
  \label{eq:1}
  C_{ij}=\sum_{l=1}^nX_{il}X_{jl}\,.
\end{equation}
In matrix notation, this can be combined as the matrix multiplication
\eq{\mat C=\mat X\mat X^T.}

Gene pairs with large positive or negative correlation values tend to
be up- or down-regulated together, due to either a direct regulatory
link between them, or being jointly co-regulated by a third, often
hidden, factor. By filtering for correlation values exceeding a
significance threshold determined by comparison with randomly permuted
data, a discrete co-expression network is obtained. Assuming that a
high degree of co-expression signifies that genes are involved in the
same biological processes, graph theoretical methods can be employed,
for instance, to predict gene function \cite{sharan2007network}.

One drawback of the Pearson correlation is that by definition it is
biased towards \emph{linear} associations. To overcome
this limitation, other measures are available. The Spearman
correlation uses expression data ranks (cf. Section
\ref{sec:genet-gene-expr}) in Equation \eqref{eq:1}, and will give high
score to \textit{monotonic} relations. Mutual information is the most
general measure and detects both linear and non-linear
associations. For a pair of discrete random variables $A$ and $B$
(representing the expression levels of two genes) taking values $a_l$
and $b_m$, respectively, the mutual information is defined as
\begin{equation*}
  MI(A,B)=H(A)+H(B)-H(A,B),
\end{equation*}
where
\begin{align*}
  H(A) &= -\sum_l P(a_l)\log P(a_l),\\
  H(B) &= -\sum_m P(b_m)\log P(b_m),\\
  H(A,B) &= \sum_{lm} P(a_l,b_m) \log P(a_l,b_m),
\end{align*}
are the individual and joint Shannon entropies of $A$ and $B$, and
$P(a_l)=P(A=a_l)$, and likewise for the other terms. Since gene
expression data are continuous, mutual information estimation is
non-trivial and usually involves some form of discretisation
\cite{daub2004}. Mutual information has been successfully used as a
co-expression measure in a variety of contexts
\cite{butte2000, basso2005, faith2007}.

\subsection{Clustering and co-expression module detection}
\label{sec:clust-co-expr}

It is generally understood that cellular functions are carried out by
``modules'', groups of molecules that operate together and whose
function is separable from that of other modules
\cite{hartwell1999}. Clustering gene expression data (i.e. dividing
genes into discrete groups on the basis of similarities in their
expression profiles) is a standard approach to detect such
functionally coherent gene modules. The literature on gene
expression clustering is vast and cannot possibly be reviewed
comprehensively here. It includes ``standard'' methods such as
hierarchical clustering \cite{eisen1998cluster}, $k$-means
\cite{tavazoie1999systematic}, graph-based methods that operate
directly on co-expression networks \cite{sharan2000click}, and
model-based clustering algorithms which assume that the data is
generated by a mixture of probability distributions, one for each
cluster \cite{medvedovic2002bayesian}. Here we briefly describe a few
recently developed methods with readily available softwares.

\paragraph{Modularity maximization} Modularity maximization is a network
clustering method that is particularly popular in the physical and
social sciences, based on the assumption that intra-module
connectivity should be much denser than inter-module connectivity
\cite{newman2004,newman2006b}. In the context of co-expression
networks, this method can be used to identify gene modules directly
from the correlation matrix $\mat C$ \cite{Ayroles:2009}.  Suppose the
genes are grouped into $N$ modules $M_l,~l=1,\dots,N$. Each module
$M_l$ is a non-empty set that can contain any combination of the genes
$i=1,\dots,k$, but each gene is contained by exactly one module.
Also define $M_0$ as the set containing all
genes. The modularity score function is defined as 
\eq{S(M)=\sum_{l=1}^N\left(\frac{W(M_l,M_l)}{W(M_0,M_0)}-\left(\frac{W(M_l,M_0)}{W(M_0,M_0)}\right)^2\right),}
where $W(A,B) =\sum_{i\in A,j\in B,i\ne j}w(C_{ij})$ is a weight
function, summing over all the edges that connect one vertex in $A$
with another vertex in $B$, and $w(x)$ is a monotonic function to map
correlation values to edge strengths. Common functions are $w(x)=|x|$,
$|x|^\beta$ (power law) \cite{Langfelder:2008}, $e^{\beta |x|}$ (exponential) \cite{Ayroles:2009}, or
$1/(1+e^{\beta x})$ (sigmoid) \cite{lee2009learning}.

A modularity maximization software particularly suited for large
networks is \textbf{Fast Modularity}
(\url{http://www.cs.unm.edu/~aaron/research/fastmodularity.htm})
\cite{Clauset:2004}.

\paragraph{Markov Cluster algorithm}

The Markov Cluster (MCL) algorithm is a graph-based clustering
algorithm, which emulates random walks among gene vertices to detect
clusters in a graph obtained directly from the co-expression matrix
$\mat C$. It is implemented in the
\textbf{MCL} software (\url{http://micans.org/mcl/}) \cite{Van-Dongen:2001,Enright:2002}. The MCL algorithm starts
with the correlation matrix $\mat C$ as the probability flow matrix of
a random walk, and then iteratively suppresses weak structures of the
network and performs a multi-step random walk. In the end, only
backbones of the network structure remain, essentially capturing
the modules of co-expression network.  To be precise, the MCL
algorithm performs the following two operations on $\mat C$
alternatingly:
\begin{itemize}
\item\textbf{Inflation:} The algorithm first contrasts stronger direct
  connections against weaker ones, using an element-wise power law
  transformation, and normalizes each column separately to sum to one,
  such that the element $C_{ij}$ corresponds to the dissipation rate
  from vertex $X_i$ to $X_j$ in a single step. The inflation operation
  hence updates $\mat C$ as
  $\mat C\rightarrow\mathbf{\Gamma}_\alpha\mat C$, where
  the contrast rate $\alpha>1$ is
  a predefined parameter of the algorithm. After operation
  $\mathbf{\Gamma}_\alpha$, each element of $\mat C$ becomes
  \eq{C_{ij}\rightarrow\mathbf{\Gamma}_\alpha
    C_{ij}=|C_{ij}|^\alpha/\sum_{p=1}^k|C_{pj}|^\alpha.}
\item\textbf{Expansion:} The probability flow matrix $\mat C$ controls
  the random walks performed in the expansion phase. After some
  integer $\beta\ge2$ steps of random walk, gene pairs with strong direct
  connections and/or strong indirect connections through other genes
  tend to see more probability flow exchanges, suggesting higher
  probabilities of belonging to the same gene modules. The expansion
  operation for the $\beta$-step random walk corresponds to the matrix
  power operation \eq{\mat C\rightarrow\mat C^\beta.}
\end{itemize}

The MCL algorithm performs the above two operations iteratively until
convergence.  Non-zero entries in the convergent matrix $\mat C$ connect
gene pairs belonging to the same cluster, whereas all inter-cluster
edges attain the value zero, so that cluster structure can be
obtained directly from this matrix
\cite{Van-Dongen:2001,Enright:2002}.

\paragraph{Weighted Gene Co-expression Network Analysis}

With higher than average correlation or edge densities within
clusters, genes from the same cluster typically share more neighbouring
(i.e. correlated) genes. The weighted number of shared neighbouring
genes hence can be another measure of gene function similarity. This
information is captured in the so-called topological overlap matrix
$\mat\Omega$, first defined in \cite{ravasz2002} for binary networks
as \eq{\omega_{ij}=\frac{A_{ij}+\sum_u
    A_{iu}A_{uj}}{\mathrm{min}(k_i,k_j)+1-A_{ij}},} where $A$ is the
(binary) adjacency matrix of the network and $k_i=\sum_uA_{iu}$ is
the connectivity of vertex $X_i$. The $\sum_uA_{iu}A_{uj}$ term
represents vertex similarity through neighbouring genes, and the rest
of terms normalise the output as $0\le\omega_{ij}\le1$.  This concept
was later extended onto networks with weighted edges by applying a
``soft threshold'' pre-process on the correlation matrix, for example
as
\begin{align*}
  A_{ij}&=\left|\frac{1+C_{ij}}{2}\right|^\alpha,\\
  \intertext{or}
  A_{ij}&=\left|C_{ij}\right|^\alpha,
\end{align*}
such that $0\le A_{ij}\le1$ \cite{zhang2005b}. Note that in the first
case only positive correlations have high edge weight, whereas in the
second case positive and negative correlations are treated
equally. The parameter $\alpha>1$ is determined such that the weighted
network with adjacency matrix $A$ has approximately a scale-free
degree distribution \cite{zhang2005b}.

In principle, any clustering algorithm (including the aforementioned
ones) can be applied to the topological overlap matrix
$\mat\Omega$. In the popular \textbf{WGCNA} software
(\url{http://labs.genetics.ucla.edu/horvath/htdocs/CoexpressionNetwork/Rpackages/WGCNA/})
\cite{Langfelder:2008}, which is a multi-purpose toolbox for network
analysis, hierachical
clustering with a dynamic tree-cut algorithm
\cite{langfelder2008defining} is employed.

\paragraph{Model-based clustering}

Model-based clustering approaches assume that the observed data is
generated by a mixture of probability distributions, one for each
cluster, and takes explicitly into account the noise of gene
expression data. To infer model parameters and cluster assignments,
techniques such as Expectation Maximization (EM) or Gibbs sampling are used
\cite{liu2002}. A recently developed method assumes that the
expression levels of genes in a cluster are random samples drawn from
a mixture of normal distributions, where each mixture component
corresponds to a clustering of samples for that module, i.e. it
performs a two-way co-clustering operation \cite{joshi2008}. The
method is available as part of the \textbf{Lemon-Tree} package
(\url{https://github.com/eb00/lemon-tree}) and has been successfully
used in a variety of applications \cite{bonnet2015}.

The co-clustering is carried out by a Gibbs sampler which iteratively
updates the assignment of each gene and, within each gene cluster, the
assignment of each experimental condition. The co-clustering operation
results the full posterior distribution, which can be written as
\begin{equation*}
  p(\mathcal{C}\mid\X) \propto \prod_{l=1}^N \prod_{u=1}^{L_l} \iint 
  p(\mu,\tau) \prod_{i\in\mathcal{M}_l}\prod_{m\in \mathcal{E}_{l,u}} p (X_{im}\mid
  \mu,\tau)\; d\mu d\tau,
\end{equation*}
where
$\mathcal{C}=\{M_l, \mathcal{E}_{l,u}\colon l=1,\dots,N;
u=1,\dots, L_l\}$
is a co-clustering consisting of $N$ gene modules $M_l$,
each of which has a set of $L_m$ sample clusters as $\mathcal{E}_{l,u}$;
$p(X_{im}\mid \mu,\tau)$ is a normal distribution function with
mean $\mu$ and precision $\tau$; and $p(\mu,\tau)$ is a
non-informative normal-gamma prior. Detailed investigations of the
convergence properties of the Gibbs sampler showed that the best
results are obtained by deriving consensus clusters from multiple
independent runs of the sampler. In the \textbf{Lemon-Tree} package, consensus
clustering is performed by a novel spectral graph clustering algorithm
\cite{michoel2012} applied to the weighted graph of
pairwise frequencies with which two genes are assigned to the same
gene module \cite{bonnet2015}.

\section{Causal gene networks}

\subsection{Using genotype data to prioritize edge directions in co-expression networks\label{sec-direction}}
Pairwise correlations between gene expression traits define undirected
co-expression networks. Several studies have shown that pairs of gene
expression traits can be causally ordered using genotype data
\cite{zhu2004, chen2007harnessing, aten2008using,
  schadt2005integrative, neto2008inferring, neto2013modeling,
  millstein2009disentangling}.  Although varying in their statistical
details, these methods conclude that gene $A$ is causal for gene $B$,
if expression of $B$ associates significantly with $A$'s eQTLs and this
association is abolished by conditioning on expression of $A$ and on
any other known confounding factors.  In essence, this is the
principle of ``Mendelian randomization'', first introduced in
epidemiology as an experimental design to detect causal effects of
environmental exposures on human health \cite{smith2003mendelian},
applied to gene expression traits.

To illustrate how these methods work, let $A$ and $B$ be two random
variables representing two gene expression traits, and let $E$ be a
random variable representing a SNP which is an eQTL for gene $A$ and
$B$. Since genotype cannot be altered by gene expression (i.e. $E$
cannot have any incoming edges), there are three possible regulatory
models to explain the joint association of $E$ to $A$ and $B$:
\begin{enumerate}
\item $E\rightarrow A\rightarrow B$: the association of $E$ to $B$ is
  indirect and due to a causal interaction from $A$ to $B$.
\item $E\rightarrow B\rightarrow A$: idem with the roles of $A$ and
  $B$ reversed.
\item $A\leftarrow E\rightarrow B$: $A$ and $B$ are independently
  associated to $E$.
\end{enumerate}
To determine if gene $A$ mediates the effect of SNP $E$ on gene $B$
(model 1), one can test whether conditioning on $A$ abolishes the
correlation between $E$ and $B$, using the partial correlation
coefficient
\begin{align*}
  \cor{E,B\mid A} =
  \frac{\cor{E,B}-\cor{E,A}\cor{B,A}}{\sqrt{(1-\cor{E,A}^2)(1-\cor{B,A}^2)}.} 
\end{align*}
If model 1 is correct, then $\cor{E,B\mid A}$ is expected to be zero,
and this can be tested for example using Fisher's $Z$ transform to assess the
significance of a sample correlation coefficient. The same approach
can be used to test model 2, and if neither is significant, it is
concluded that no inference on the causal direction between $A$ and
$B$ can be made (using SNP $E$), i.e. that model 3 is correct. For
more details, see \cite{aten2008using}, who have implemented this
approach in the \textbf{NEO} software
(\url{http://labs.genetics.ucla.edu/horvath/htdocs/aten/NEO/}).

Other approaches are based on the same principle, but use statistical
model selection to identify the most likely causal model, with
the probability density functions (PDF) for the models below: 
\begin{itemize}
\item $p(E,A,B)=p(E)p(A\mid E)p(B\mid A)$,
\item $p(E,A,B)=p(E)p(B\mid E)p(A\mid B)$,
\item $p(E,A,B)=p(E)p(A\mid E) p(B\mid E,A)$,
\end{itemize}
where the dependence on $A$ in the last term of the last model indicates that
there may be a residual correlation between $B$ and $A$ not explained
by $E$. The minimal additive model assumes the distributions are
\cite{schadt2005integrative} \eqa{
  E&\sim&\mathrm{Bernoulli}(q),\nonumber\\
  A\mid E&\sim&\mathrm{N}(\mu_{A\mid E},\sigma_A^2),\nonumber\\
  B\mid A&\sim&\mathrm{N}\left(\mu_B+\rho\frac{\sigma_B}{\sigma_A}(A-\mu_A),(1-\rho^2)\sigma_B^2\right),\nonumber\\
  B\mid E,A&\sim&\mathrm{N}\left(\mu_{B\mid E}+\rho\frac{\sigma_B}{\sigma_A}(A-\mu_{A\mid E}),(1-\rho^2)\sigma_B^2\right),\nonumber}
so that $E$ fulfills a Bernoulli distribution, $A\mid E$ undergoes a
normal distribution whose mean depends on $E$, and that $B\mid A$ has
a conditional normal distribution whose mean and variance are contributed
in part by $A$. For $(B\mid E,A)$, the mean of $B$ also
depends on $E$. The parameters of all distributions can be estimated by
maximum likelihood, and the model with the
highest likelihood is selected as the most likely causal model.
The number of free parameters can be accounted using
penalties like the Akaike information criterion (AIC)
\cite{schadt2005integrative}.

The approach has been extended in various ways. In
\cite{chen2007harnessing}, likelihood ratio tests, comparison to
randomly permuted data, and false discovery rate estimation techniques
are used to convert the three model scores in a single probability
value $P(A\to B)$ for a causal interaction from gene $A$ to $B$. This
method is available in the \textbf{Trigger} software
(\url{https://www.bioconductor.org/packages/release/bioc/html/trigger.html}). In
\cite{millstein2009disentangling} and \cite{neto2013modeling}, the
model selection task is recast into a single hypothesis test, using
$F$-tests and Vuong's model selection test respectively, resulting in
a significance $p$-value for each gene-gene causal interaction.

It should be noted that all of the above approaches suffer from
limitations due to their inherent model assumptions. In particular, the
presence of unequal levels of measurement noise among genes, or of
hidden regulatory factors causing additional correlation among genes,
can confuse causal inference. For example, excessive error level in
the expression data of gene $A$, may mistake the true structure
$E\rightarrow A\rightarrow B$ as $E\rightarrow B\rightarrow A$. These
limitations are discussed in
\cite{rockman2008reverse,li2010critical}.

\subsection{Using Bayesian networks to identify causal regulatory mechanisms}

Bayesian networks are probabilistic graphical models which encode
conditional dependencies between random variables in a directed
acyclic graph (DAG). Although Bayesian network cannot fully reflect
certain pathways in gene regulation, such as self-regulation or
feedback loops, they still serve as a popular method for modelling
gene regulation networks, as they provide a clear methodology for
learning statistical dependency structures from possibly noisy data
\cite{Friedman:1999,Friedman:2000,koller2009}.

We adopt our previous convention in Section \ref{sec:genet-gene-expr},
where we have the gene expression data $\mat X$ and genetic markers
$\mat G$. The model contains a total of $k$ vertices (i.e. random
variables), $X_i$ with $i=1,\dots,k$, corresponding to the expression
level of gene $i$. Given a DAG $\G$, and denoting the parental vertex
set of $X_i$ by $\mat{Pa}^{(\G)}(X_i)$, the acyclic property of $\G$ allows
to define the joint probability distribution function as
\begin{equation}
  \label{eq:2}
  p(X_1,\dots,X_k\mid\G)
  =\prod_{i=1}^kp(X_i \mid \mat{Pa}^{(\G)}(X_i)).
\end{equation}
In its simplest form, we model the conditional distributions as
\begin{equation*}
  p\bigl(X_i\mid \mat{Pa}^{(\G)}(X_i)\bigl) = N\biggl(\alpha_i +
  \sum_{X_j\in\mat{Pa}^{(\G)}(X_i)} \beta_{ji}(X_j-\alpha_j),\sigma_i^2\biggr),
\end{equation*}
where $(\alpha_i,\sigma_i)$ and $\beta_{ji}$ are parameters for vertex $X_i$
and edge $X_j\rightarrow X_i$ respectively, as part of the DAG structure $\G$.
Under such modelling, the Bayesian network is called a linear Gaussian
network.

The likelihood of data $\mat X$ given the graph $\G$ is
\begin{align*}
  p(\mat X\mid\G) =  \prod_{i=1}^k\prod_{l=1}^n 
  p(X_{il}\mid \{X_{jl}, X_j\in\mat{Pa}^{(\G)}(X_i)\}).
\end{align*}
Using Bayes' rule, the log-likelihood of the DAG $\G$ based on the
gene expression data $\mat X$ becomes \eq{\log p(\G\mid\mat X)=\log
  p(\mat X\mid \G)+\log p(\G)-\log p(\mat X),} where $p(\G)$ is the
prior probability for $\G$, and $p(\mat X)$ is a constant when the expression
data is provided, so the follow-up calculations do not rely on it.

Typically, a locally optimal DAG is found by starting from a random graph and randomly
ascending the likelihood by adding, modifying, or removing one directed edge at a time
\cite{Friedman:1999,Friedman:2000,koller2009}.
Alternatively, the posterior distribution $p(\G\mid\mat X)$ can be estimated with
Bayesian inference using Markov Chain Monte Carlo (MCMC) simulation, allowing
us to estimate the significance levels at an extra computational cost.
The parameter values of $\alpha$, $\beta$, and $\sigma$, as part of $\G$, can be estimated
with maximum likelihood.

When Bayesian network is modified by a single edge, only the vertices that
receive a change would require a recalculation, whilst all others
remain intact. This significantly reduces the amount of computation
needed for each random step. A further speedup is achievable if we
constrain the maximum number of parents each vertex can have, either
by using the same fixed number for all nodes, or by
pre-selecting a variable number of potential parents for each node
using, for instance, a preliminary $L_1$-regularisation step
\cite{schmidt2007learning}.

Two DAGs are called Markov equivalent if they result in the same PDF
\cite{koller2009}.  Clearly, using gene expression data
alone, Bayesian networks can only be resolved up to Markov
equivalence. To break this equivalence and uncover a more specific
causal gene regulation network, genotype data is incorporated in the
model inference process. The most straightforward approach is to use
any of the methods in the previous section to calculate the
probability $P(X_i\to X_j)$ of a causal interaction from $X_i$ to $X_j$
\cite{zhu2004,zhu2008,zhu2012stitching, zhang2013integrated}, for example
by defining the prior as
$p(\G)=\prod_{X_i}\left(\prod_{X_j\in\mat{Pa}^{(\G)}(X_i)}P(X_j\to X_i)
\prod_{X_j\not\in\mat{Pa}^{(\G)}(X_i)}(1-P(X_j\to X_i))\right)$.
A more
ambitious approach is to jointly learn the eQTL associations and causal
trait (i.e. gene or phenotype) networks. In \cite{neto2010causal},
EM is used to alternatingly map eQTLs given the
current DAG structure, and update the DAG structure and model
parameters given the current eQTL mapping. In
\cite{scutari2014multiple}, Bayesian networks are learned where SNPs
and traits both enter as variables in the model, with the constraint
that traits can depend on SNPs, but not vice versa. However, the
additional complexity of both methods means that they are
computationally expensive and have only been applied to problems with
a handful of traits \cite{neto2010causal,scutari2014multiple}.

A few additional ``tips and tricks'' are worth mentioning:
\begin{itemize}
\item First, when
the number of vertices is much larger than the sample count, we may
break the problem into independent sub-problems by learning a separate
Bayesian network for each co-expression module
(Section \ref{sec-coex} and \cite{zhang2013integrated}).
Dependencies between modules could then
be learned as a Bayesian network among the module eigengenes
\cite{langfelder2007eigengene}, although this does not seem to have
been explored.
\item Second, Bayesian network learning algorithms
inevitably result in locally optimal models which may contain a high
number of false positives. To address this problem, we can
run the algorithm multiple times and report an averaged network,
only consisting of edges which appear sufficiently frequent.
\item Finally, another technique that helps in
distinguishing genuine dependencies from false positives is
\emph{bootstrapping}, where resampling with replacement is executed on
the existing sample pool. A fixed number of samples
are randomly selected and then
processed to predict a Bayesian network. This process is
repeated many times, essentially regarding the distribution of sample pool
as the true PDF, and allowing to estimate the robustness of each
predicted edge, so that only those with high significance are retained
\cite{friedman1999data}. In theory, even the whole pipeline of Figure
\ref{fig-flow} up to the \textit{in silico} validation could be
simulated in this way. Although bootstrapping is computationally
expensive and mostly suited for small datasets, it could be used in
conjunction with the separation into modules on larger datasets.
\end{itemize}

\subsection{Using module networks to identify causal regulatory mechanisms}

Module network inference is a statistically well-grounded method which
uses probabilistic graphical models to reconstruct modules of
co-regulated genes and their upstream regulatory programs, and which
has been proven useful in many biological case studies
\cite{add1,segal2003,friedman2004,bonnet2015}. The module network
model was originally introduced as a method to infer regulatory
networks from large-scale gene expression compendia, as implemented in
the \textbf{Genomica} software
(\url{http://genomica.weizmann.ac.il})\cite{segal2003}.  Subsequently
the method has been extended to integrate eQTL and gene expression
data \cite{lee2006, lee2009learning, zhang2010bayesian}.  The module
network model starts from the same formula as Equation
\eqref{eq:2}. It is then assumed that genes belonging to the same
module share the same parents and conditional distributions; these
conditional distributions are parameterized as decision trees, with
the parental genes on the internal (decision) nodes and normal
distributions on the leaf nodes \cite{segal2003}. Recent algorithmic
innovations decouple the module assignment and tree structure learning
from the parental gene assignment and use Gibbs sampling and ensemble
methods for improved module network inference \cite{joshi2008,
  joshi2009}. These algorithms are implemented in the
\textbf{Lemon-Tree} software
(\url{https://github.com/eb00/lemon-tree}), a command line software
suite for module network inference \cite{bonnet2015}.

\subsection{Illustrative example}
\label{sec:illustrative-example}

\begin{figure}[tbp]
\centering
\includegraphics{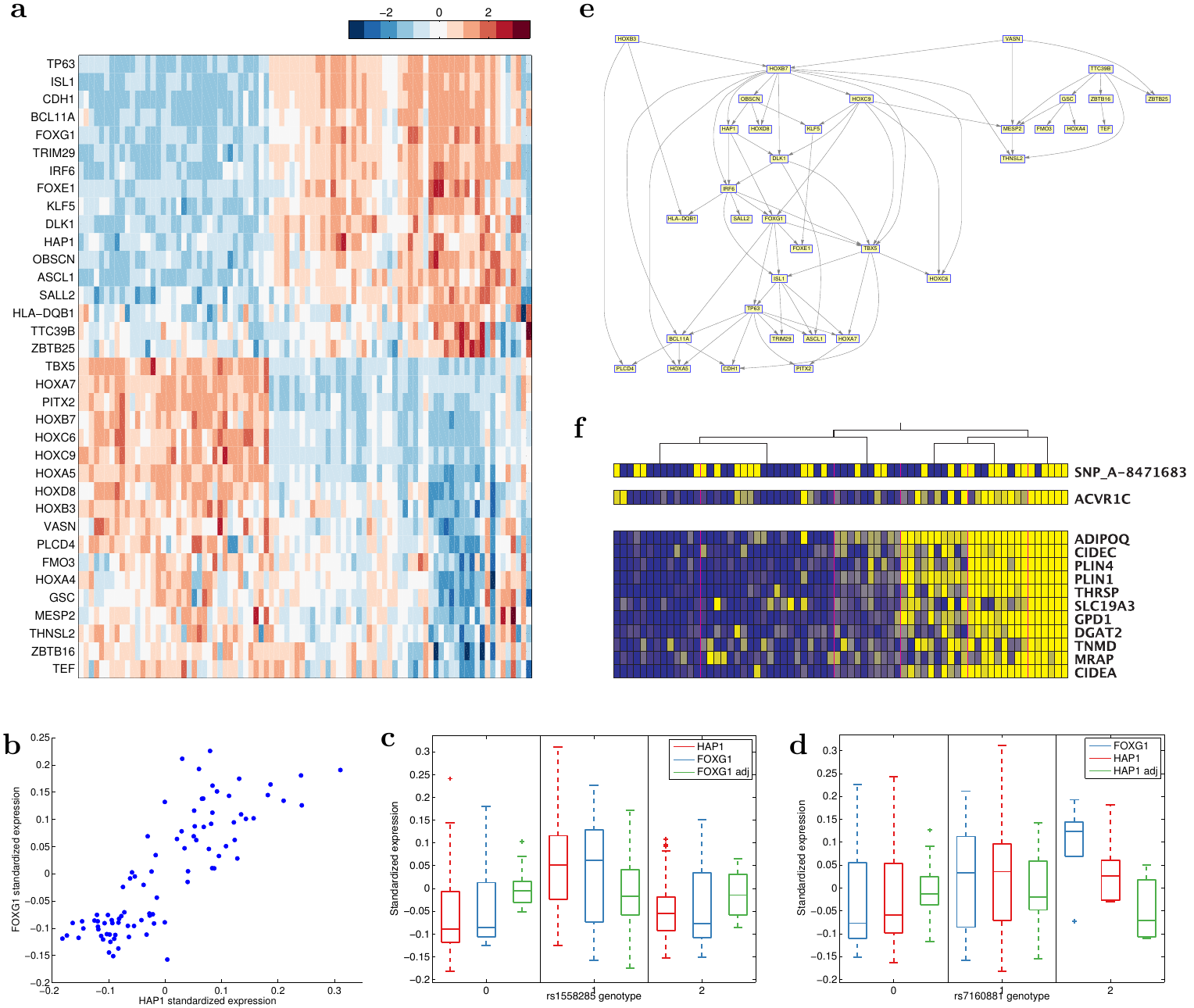}
\caption{
\textbf{(a)} Heatmap of standardized expression profiles across
88 visceral fat samples for 10 eQTL genes and
25 TFs belonging to a co-expression cluster inferred from the
STAGE data.
\textbf{(b)} Co-expression of HAP1 and FOXG1 across
88 visceral fat samples.
\textbf{(c)} Association between HAP1's eQTL (rs1558285) and
expression of HAP1 (red), FOXG1 (blue) and FOXG1 adjusted for HAP1
and FOXG1's eQTL (green).
\textbf{(d)} Association between FOXG1's eQTL (rs7160881) and expression of
FOXG1 (blue), HAP1 (red), and HAP1 adjusted for FOXG1
and HAP1's eQTL (green).
\textbf{(e)} Causal interactions
inferred between the same genes as in (a) using Bayesian network
inference.
\textbf{(f)} Example of a
regulatory module inferred by \textbf{Lemon-Tree} from the STAGE
data.
See Section \ref{sec:illustrative-example} for further details.\label{fig-figs}}
\end{figure}

We have recently identified genome-wide significant eQTLs for 6,500
genes in seven tissues from the Stockholm Atherosclerosis Gene
Expression (STAGE) study \cite{foroughi2015}, and performed
co-expression clustering and causal networks reconstruction
\cite{talukdar2015}. To illustrate the above concepts, we show some
results for a co-expression cluster in visceral fat (88 samples, 324
genes) which was highly enriched for tissue development genes
($P=5\times 10^{-10}$) and contained 10 genome-wide significant eQTL
genes and 25 transcription factors, including eight members of the
homeobox family (Figure \ref{fig-figs}a).

A representative example of an inferred causal interaction is given by
the co-expression interaction between HAP1 (huntingtin-associated
protein 1, chr17 q21.2-21.3) and FOXG1 (forkhead box G1, chr14
q11-q13). The expression of both genes is highly correlated
($\rho=0.85$, $P=4.4\times 10^{-24}$, Figure \ref{fig-figs}b). HAP1
expression shows a significant, non-linear association with its eQTL
rs1558285 ($P=1.2\times 10^{-4}$); this SNP also associates
significantly with FOXG1 expression in the cross-association test
($P=0.0024$), but not anymore after conditioning FOXG1 on HAP1 and its
own eQTL rs7160881 ($P=0.67$) (Figure \ref{fig-figs}c). In contrast,
although FOXG1 expression is significantly associated with its eQTL
rs7160881 ($P=0.0028$), there is no association between this SNP and
HAP1 expression ($P=0.037$), and conditioning on FOXG1 and HAP1's eQTL
has only a limited effect ($P=0.19$)
(Figure \ref{fig-figs}d). Using conditional independence tests
(Section \ref{sec-direction}), this results in a high-confidence
prediction that HAP1 $\to$ FOXG1 is causal.

A standard greedy Bayesian network search algorithm
\cite{schmidt2007learning} was run on the aforementioned cluster of
324 genes. Figure \ref{fig-figs}e shows the predicted consensus
sub-network of causal interactions between the 10 eQTLs and 25
TFs. This illustrates how a sparse Bayesian network can accurately
represent the fully connected co-expression network (all 35 genes have
high-mutual co-expression, cf. Figure \ref{fig-figs}a).

Figure \ref{fig-figs}f shows a typical regulatory module inferred by the
\textbf{Lemon-Tree} software, also from the STAGE data. Here a heatmap is shown
of the genotypes of an eQTL (top), the expression levels of a
regulatory gene (middle), predicted to regulate a co-expression module
of 11 genes (bottom). The red lines indicate sample clusters
representing separate normal distributions inferred by the model-based
co-clustering algorithm (Section \ref{sec:clust-co-expr}).

\section{\textit{In silico} validation of predicted gene regulation
  networks}
\label{sec:silico-valid-pred}

Gene regulation networks reconstructed from omics data represent hypotheses about
the downstream molecular implications of genetic variations in a
particular cell or tissue type. An essential first step towards using
these networks in concrete applications (e.g. discovering novel
candidate drug target genes and pathways) consists of validating them
using independent data. The following is a non-exhaustive list of
typical \textit{in silico} validation experiments.

\paragraph{Model likelihood comparison and cross-validation.}
When different algorithms are used to infer gene
network models, their log-likelihoods can be compared to select the
best one. (With the caveat that the same data that was used to learn
the models is used to compare them, this comparison is
meaningful only when the algorithms optimize \emph{exactly} the same
(penalized) log-likelihood functions.) In a $K$-fold cross-validation
experiment, the available samples are divided into $K$ subsets of
approximately equal size. For each subset, models are learned from a
dataset consisting of the $K-1$ other subsets, and the model
likelihood is calculated using only the unseen data subset. Thus,
cross-validation is used to test the generalisability of the inferred
network models to unseen data. For an example where model likelihood
comparison and cross-validation were used to compare two module
network inference strategies, see \cite{joshi2009}.

\paragraph{Functional enrichment.} Organism-specific gene ontology
databases contain structured functional gene annotations
\cite{ashb00}. These databases can be used to construct gene signature
sets composed of genes annotated to the same biological process,
molecular function or cellular component. Reconstructed gene networks
can then be validated by testing for enriched connectivity of gene
signature sets using a method proposed by \cite{zhu2008}. For a given
gene set, this method considers all network nodes belonging to the set
and their nearest neighbours, and from this set of nodes and edges,
the largest connected sub-network is identified. Then the
enrichment of the gene set in this sub-network is tested using the
Fisher exact test and compared to the enrichment of randomly selected
gene sets of the same size.

\paragraph{Comparison with physical interaction networks.} Networks of
transcription factor - target interactions based on ChIP-sequencing
data \cite{furey2012chip} from diverse cell and tissue types are
available from the \textbf{ENCODE} \cite{encode2012}, \textbf{Roadmap Epigenomics}
\cite{kundaje2015integrative} and \textbf{modENCODE}
\cite{gerstein2010integrative, roy2010identification,
  yue2014comparative} projects, while physical 
protein-protein interaction networks are available for many organisms
through databases such as the \textbf{BioGRID} \cite{chatr2014biogrid}. Due to
indirect effects, networks predicted from gene expression data rarely
show a significant overlap with networks of direct physical
interactions. A more appropriate validation is therefore to test for
enrichment for short connection paths in the physical networks between
pairs predicted to interact in the reconstructed networks
\cite{bonnet2015}.

\paragraph{Gene perturbation experiments.} 
Gene knock-out experiments provide the ultimate gold standard of a
causal network intervention, and genes differentially expressed
between knock-out and control experiments can be considered as true
positive direct or indirect targets of the knocked-out gene.
Predicted gene networks can be validated by compiling relevant
(i.e. performed in a relevant cell or tissue type) gene knock-out
experiments from the \textbf{Gene Expression Omnibus}
(\url{http://www.ncbi.nlm.nih.gov/geo/}) or \textbf{ArrayExpress}
(\url{https://www.ebi.ac.uk/arrayexpress/}) and comparing the overlap
between gene sets responding to a gene knock-out and network genes
predicted to be downstream of the knocked-out gene. Overlap
significance can be estimated by using randomized networks with the
same degree distribution as the predicted network.

\section{Future perspective: Integration of multi-omics data}
 
Although combining genotype and transcriptome data to reconstruct
causal gene networks has led to important discoveries in a variety of
applications \cite{civelek2014systems}, important details are not
incorporated in the resulting network models, particularly regarding
the causal molecular mechanisms linking eQTLs to their target genes,
and the relation between variation in transcript levels and protein
levels, with the latter ultimately determining phenotypic
responses. Several recent studies have shown that at the molecular
level, \textit{cis}-eQTLs primarily cause variation in transcription
factor binding to gene regulatory DNA elements, which then causes
changes in histone modifications, DNA methylation and mRNA expression
of nearby genes (reviewed in \cite{albert2015role}). Although mRNA
expression can be used as a surrogate for protein expression, due to
diverse post-transcriptional regulation mechanisms,
the correlation between mRNA and protein levels is known to be modest
\cite{lu2007,schwanhausser2011}, and genetic loci that affect mRNA and
protein expression levels do not always overlap
\cite{foss2007,wu2013variation}. Thus, an ideal systems genetics study
would integrate genotype data and molecular measurements at all levels
of gene regulation from a large number of individuals.

Human lymphblastoid cell lines (LCL) are emerging as the primary model
system to test such a approach. Whole-genome mRNA and micro-RNA
sequencing data are available for 462 LCL samples from five
populations genotyped by the 1000 Genomes Project
\cite{lappalainen2013transcriptome}; protein levels from quantitative
mass spectrometry for 95 samples \cite{wu2013variation}; ribosome
occupancy levels from sequencing of ribosome-protected mRNA for 50
samples \cite{cenik2015integrative}; DNA-occupancy levels of the
regulatory TF PU.1, the RNA polymerase II subunit RBP2, and three
histone modifications from ChIP-sequencing of 47 samples
\cite{waszak2015population}; and the same three histone modifications
from ChIP-sequencing of 75 samples \cite{grubert2015genetic}. These
population-level datasets can be combined further with
three-dimensional chromatin contact data from Hi-C \cite{rao20143d}
and ChIA-PET \cite{grubert2015genetic}, knock-down experiments
followed by microarray measurements for 59 transcription-associated
factors and chromatin modifiers \cite{cusanovich2014functional}, as
well as more than 260 ENCODE assays (including ChIP-sequencing of 130
TFs) \cite{encode2012} in a reference LCL cell line (GM12878).
Although the number of samples where all measures are simultaneously
available is currently small, this number is sure to rise in the
coming years, along with the availability of similar measurements in
other cell types. Despite the challenging heterogeneity of data and
analyses in the integration of multi-omics data, web-based toolboxes,
such as \textbf{GenomeSpace} (\url{http://www.genomespace.org})
\cite{add1} can prove helpful to non-programmer researchers.

\section{Conclusions}
\label{sec:conclusions}

In this chapter we have reviewed the main methods and softwares to
carry out a systems genetics analysis, which combines genotype and
various omics data to identify eQTLs and their associated genes,
reconstruct co-expression networks and modules, reconstruct causal
Bayesian gene and module networks, and validate predicted networks
\textit{in silico}. Several method and software options are available
for each of these steps, and by necessity a subjective choice about
which ones to include had to be made, based largely on their ability
to handle large datasets, their popularity in the field, and our
personal experience of using them. Where methods have been compared in
the literature, they have usually been performed on a small number of
datasets for a specific subset of tasks, and results have rarely been
conclusive. That is, although each of the presented methods will give
somewhat different results, no objective measurements will
consistently select one of them as the ``best'' one. Given this lack of
objective criterion, the reader may well prefer to use a single
software that allows to perform all of the presented analyses, but
such an integrated software does not currently exist.

Nearly all of the examples discussed referred to the integration of
genotype and transcriptome data, reflecting the current dominant
availability of these two data types. However, omics technologies are
evolving at a fast pace, and it is clear that data on the variation of
TF binding, histone modifications, and post-transcriptional and
protein expression levels will soon become more widely available.
Developing appropriate statistical models and computational methods to
infer causal gene regulation networks from these multi-omics datasets
is surely the most important challenge for the field.

\section*{Acknowledgements}

The authors' work is supported by the BBSRC [BB/M020053/1] and Roslin
Institute Strategic Grant funding from the BBSRC [BB/J004235/1].


\begin{thebibliography}{100}

\bibitem{mackay2009genetics}
Mackay TF, Stone EA and Ayroles JF.
\newblock The genetics of quantitative traits: challenges and prospects.
\newblock \emph{Nature Reviews Genetics} \textbf{10}:565--577 (2009).

\bibitem{goddard2009mapping}
Goddard ME and Hayes BJ.
\newblock Mapping genes for complex traits in domestic animals and their use in
  breeding programmes.
\newblock \emph{Nature Reviews Genetics} \textbf{10}:381--391 (2009).

\bibitem{manolio2013bringing}
Manolio TA.
\newblock Bringing genome-wide association findings into clinical use.
\newblock \emph{Nature Reviews Genetics} \textbf{14}:549--558 (2013).

\bibitem{hindorff2009potential}
Hindorff LA \emph{et~al.}
\newblock Potential etiologic and functional implications of genome-wide
  association loci for human diseases and traits.
\newblock \emph{Proceedings of the National Academy of Sciences}
  \textbf{106}:9362--9367 (2009).

\bibitem{schaub2012linking}
Schaub MA \emph{et~al.}
\newblock Linking disease associations with regulatory information in the human
  genome.
\newblock \emph{Genome Research} \textbf{22}:1748--1759 (2012).

\bibitem{rockman2006genetics}
Rockman MV and Kruglyak L.
\newblock Genetics of global gene expression.
\newblock \emph{Nature Reviews Genetics} \textbf{7}:862--872 (2006).

\bibitem{georges2007mapping}
Georges M.
\newblock Mapping, fine mapping, and molecular dissection of quantitative trait
  loci in domestic animals.
\newblock \emph{Annu Rev Genomics Hum Genet} \textbf{8}:131--162 (2007).

\bibitem{cookson2009mapping}
Cookson W \emph{et~al.}
\newblock Mapping complex disease traits with global gene expression.
\newblock \emph{Nature Reviews Genetics} \textbf{10}:184--194 (2009).

\bibitem{cheung2009genetics}
Cheung VG and Spielman RS.
\newblock Genetics of human gene expression: mapping dna variants that
  influence gene expression.
\newblock \emph{Nature Reviews Genetics} \textbf{10}:595--604 (2009).

\bibitem{cubillos2012lessons}
Cubillos FA, Coustham V and Loudet O.
\newblock Lessons from {eQTL} mapping studies: non-coding regions and their
  role behind natural phenotypic variation in plants.
\newblock \emph{Current Opinion in Plant Biology} \textbf{15}:192--198 (2012).

\bibitem{dimas2009common}
Dimas AS \emph{et~al.}
\newblock Common regulatory variation impacts gene expression in a cell
  type--dependent manner.
\newblock \emph{Science} \textbf{325}:1246--1250 (2009).

\bibitem{foroughi2015}
Foroughi~Asl H \emph{et~al.}
\newblock Expression quantitative trait loci acting across multiple tissues are
  enriched in inherited risk of coronary artery disease.
\newblock \emph{Circulation: Cardiovascular Genetics} \textbf{8}:305--315
  (2015).

\bibitem{greenawalt2011survey}
Greenawalt DM \emph{et~al.}
\newblock A survey of the genetics of stomach, liver, and adipose gene
  expression from a morbidly obese cohort.
\newblock \emph{Genome Research} \textbf{21}:1008--1016 (2011).

\bibitem{ardlie2015genotype}
Ardlie KG \emph{et~al.}
\newblock The genotype-tissue expression ({GTEx}) pilot analysis: Multitissue
  gene regulation in humans.
\newblock \emph{Science} \textbf{348}:648--660 (2015).

\bibitem{albert2015role}
Albert FW and Kruglyak L.
\newblock The role of regulatory variation in complex traits and disease.
\newblock \emph{Nature Reviews Genetics} \textbf{16}:197--212 (2015).

\bibitem{williams2006expression}
Williams RW.
\newblock Expression genetics and the phenotype revolution.
\newblock \emph{Mammalian Genome} \textbf{17}:496--502 (2006).

\bibitem{kadarmideen2006genetical}
Kadarmideen HN, von Rohr P and Janss LL.
\newblock From genetical genomics to systems genetics: potential applications
  in quantitative genomics and animal breeding.
\newblock \emph{Mammalian Genome} \textbf{17}:548--564 (2006).

\bibitem{rockman2008reverse}
Rockman MV.
\newblock Reverse engineering the genotype--phenotype map with natural genetic
  variation.
\newblock \emph{Nature} \textbf{456}:738--744 (2008).

\bibitem{schadt2009}
Schadt EE.
\newblock Molecular networks as sensors and drivers of common human diseases.
\newblock \emph{Nature} \textbf{461}:218--223 (2009).

\bibitem{schadt2012new}
Schadt EE and Bj{\"o}rkegren JL.
\newblock New: network-enabled wisdom in biology, medicine, and health care.
\newblock \emph{Science Translational Medicine} \textbf{4}:115rv1--115rv1
  (2012).

\bibitem{civelek2014systems}
Civelek M and Lusis AJ.
\newblock Systems genetics approaches to understand complex traits.
\newblock \emph{Nature Reviews Genetics} \textbf{15}:34--48 (2014).

\bibitem{bjorkegren2015genome}
Bj{\"o}rkegren JL \emph{et~al.}
\newblock Genome-wide significant loci: How important are they?: Systems
  genetics to understand heritability of coronary artery disease and other
  common complex disorders.
\newblock \emph{Journal of the American College of Cardiology}
  \textbf{65}:830--845 (2015).

\bibitem{chen2008}
Chen Y \emph{et~al.}
\newblock Variations in {DNA} elucidate molecular networks that cause disease.
\newblock \emph{Nature} \textbf{452}:429--435 (2008).

\bibitem{schadt2009b}
Schadt EE, Friend SH and Shaywitz DA.
\newblock A network view of disease and compound screening.
\newblock \emph{Nat Rev Drug Disc} \textbf{8}:286--295 (2009).

\bibitem{walhout2006unraveling}
Walhout AJ.
\newblock Unraveling transcription regulatory networks by protein--{DNA} and
  protein--protein interaction mapping.
\newblock \emph{Genome Research} \textbf{16}:1445--1454 (2006).

\bibitem{ritchie2015methods}
Ritchie MD \emph{et~al.}
\newblock Methods of integrating data to uncover genotype-phenotype
  interactions.
\newblock \emph{Nature Reviews Genetics} \textbf{16}:85--97 (2015).

\bibitem{stegle2012using}
Stegle O \emph{et~al.}
\newblock Using probabilistic estimation of expression residuals (peer) to
  obtain increased power and interpretability of gene expression analyses.
\newblock \emph{Nature Protocols} \textbf{7}:500--507 (2012).

\bibitem{foss2007}
Foss EJ \emph{et~al.}
\newblock Genetic basis of proteome variation in yeast.
\newblock \emph{Nature Genetics} \textbf{39}:1369--1375 (2007).

\bibitem{nicholson2011genome}
Nicholson G \emph{et~al.}
\newblock A genome-wide metabolic {QTL} analysis in {E}uropeans implicates two
  loci shaped by recent positive selection.
\newblock \emph{PLoS Genetics} \textbf{7}:e1002270 (2011).

\bibitem{laird2011}
Laird N and Lange C.
\newblock \emph{The Fundamentals of Modern Statistical Genetics}
  (Springer2011).

\bibitem{shabalin2012matrix}
Shabalin AA.
\newblock Matrix {eQTL}: ultra fast {eQTL} analysis via large matrix
  operations.
\newblock \emph{Bioinformatics} \textbf{28}:1353--1358 (2012).

\bibitem{qi2014}
Qi J \emph{et~al.}
\newblock {kruX}: Matrix-based non-parametric {eQTL} discovery.
\newblock \emph{BMC Bioinformatics} \textbf{15}:11 (2014).

\bibitem{brem2002}
Brem RB \emph{et~al.}
\newblock Genetic dissection of transcriptional regulation in budding yeast.
\newblock \emph{Science} \textbf{296}:752--755 (2002).

\bibitem{schadt2008}
Schadt EE \emph{et~al.}
\newblock {{M}apping the genetic architecture of gene expression in human
  liver}.
\newblock \emph{PLoS Biol} \textbf{6}:e107 (2008).

\bibitem{lappalainen2013transcriptome}
Lappalainen T \emph{et~al.}
\newblock Transcriptome and genome sequencing uncovers functional variation in
  humans.
\newblock \emph{Nature} \textbf{501}:506--511 (2013).

\bibitem{golub1996}
Golub GH and Van~Loan CF.
\newblock \emph{Matrix computations}  (The Johns Hopkins University Press1996),
  third edn.

\bibitem{hemani2014detection}
Hemani G \emph{et~al.}
\newblock Detection and replication of epistasis influencing transcription in
  humans.
\newblock \emph{Nature} \textbf{508}:249--253 (2014).

\bibitem{sharan2007network}
Sharan R, Ulitsky I and Shamir R.
\newblock Network-based prediction of protein function.
\newblock \emph{Molecular Systems Biology} \textbf{3}:88 (2007).

\bibitem{daub2004}
Daub CO \emph{et~al.}
\newblock Estimating mutual information using {B}-spline functions -- an
  improved similarity measure for analysing gene expression data.
\newblock \emph{BMC Bioinformatics} \textbf{5}:118 (2004).

\bibitem{butte2000}
Butte A and Kohane I.
\newblock Mutual information relevance networks: Functional genomic clustering
  using pairwise entropy measurements.
\newblock \emph{Pac Symp Biocomputing} \textbf{5}:415--426 (2000).

\bibitem{basso2005}
Basso K \emph{et~al.}
\newblock Reverse engineering of regulatory networks in human b cells.
\newblock \emph{Nat Genet} \textbf{37}:382--390 (2005).

\bibitem{faith2007}
Faith JJ \emph{et~al.}
\newblock Large-scale mapping and validation of \textit{Escherichia coli}
  transcriptional regulation from a compendium of expression profiles.
\newblock \emph{PLoS Biol} \textbf{5}:e8 (2007).

\bibitem{hartwell1999}
Hartwell LH \emph{et~al.}
\newblock From molecular to modular cell biology.
\newblock \emph{Nature} \textbf{402}:C47--C52 (1999).

\bibitem{eisen1998cluster}
Eisen MB \emph{et~al.}
\newblock Cluster analysis and display of genome-wide expression patterns.
\newblock \emph{PNAS} \textbf{95}:14863--14868 (1998).

\bibitem{tavazoie1999systematic}
Tavazoie S \emph{et~al.}
\newblock Systematic determination of genetic network architecture.
\newblock \emph{Nature Genetics} \textbf{22}:281--285 (1999).

\bibitem{sharan2000click}
Sharan R and Shamir R.
\newblock {CLICK}: a clustering algorithm with applications to gene expression
  analysis.
\newblock In \emph{Proc Int Conf Intell Syst Mol Biol}, vol.~8, 16 (2000).

\bibitem{medvedovic2002bayesian}
Medvedovic M and Sivaganesan S.
\newblock Bayesian infinite mixture model based clustering of gene expression
  profiles.
\newblock \emph{Bioinformatics} \textbf{18}:1194--1206 (2002).

\bibitem{newman2004}
Newman MEJ and Girvan M.
\newblock Finding and evaluating community structure in networks.
\newblock \emph{Phys Rev E} \textbf{69}:026113 (2004).

\bibitem{newman2006b}
Newman MEJ.
\newblock Modularity and community structure in networks.
\newblock \emph{PNAS} \textbf{103}:8577--8582 (2006).

\bibitem{Ayroles:2009}
Ayroles JF \emph{et~al.}
\newblock Systems genetics of complex traits in drosophila melanogaster.
\newblock \emph{Nat Genet} \textbf{41}:299--307 (2009).

\bibitem{Langfelder:2008}
Langfelder P and Horvath S.
\newblock Wgcna: an r package for weighted correlation network analysis.
\newblock \emph{BMC Bioinformatics} \textbf{9}:559 (2008).

\bibitem{lee2009learning}
Lee SI \emph{et~al.}
\newblock Learning a prior on regulatory potential from eqtl data.
\newblock \emph{PLoS Genetics} \textbf{5}:e1000358 (2009).

\bibitem{Clauset:2004}
Clauset A, Newman MEJ and Moore C.
\newblock Finding community structure in very large networks.
\newblock \emph{Phys Rev E} \textbf{70}:066111 (2004).

\bibitem{Van-Dongen:2001}
Van~Dongen SM.
\newblock Graph clustering by flow simulation  (2001).

\bibitem{Enright:2002}
Enright AJ, Van~Dongen S and Ouzounis CA.
\newblock An efficient algorithm for large-scale detection of protein families.
\newblock \emph{Nucleic Acids Research} \textbf{30}:1575--1584 (2002).

\bibitem{ravasz2002}
Ravasz E \emph{et~al.}
\newblock Hierarchical organization of modularity in metabolic networks.
\newblock \emph{Science} \textbf{297}:1551--1555 (2002).

\bibitem{zhang2005b}
Zhang B and Horvath S.
\newblock A general framework for weighted gene co-expression network analysis.
\newblock \emph{Stat Appl Genet Mol Biol} \textbf{4}:17 (2005).

\bibitem{langfelder2008defining}
Langfelder P, Zhang B and Horvath S.
\newblock Defining clusters from a hierarchical cluster tree: the dynamic tree
  cut package for r.
\newblock \emph{Bioinformatics} \textbf{24}:719--720 (2008).

\bibitem{liu2002}
Liu JS.
\newblock \emph{{M}onte {C}arlo strategies in scientific computing}
  (Springer2002).

\bibitem{joshi2008}
Joshi A, Van~de Peer Y and Michoel T.
\newblock Analysis of a {Gibbs} sampler for model based clustering of gene
  expression data.
\newblock \emph{Bioinformatics} \textbf{24}:176--183 (2008).

\bibitem{bonnet2015}
Bonnet E, Calzone L and Michoel T.
\newblock Integrative multi-omics module network inference with {Lemon-Tree}.
\newblock \emph{PLoS Computational Biology} \textbf{11} (2015).

\bibitem{michoel2012}
Michoel T and Nachtergaele B.
\newblock Alignment and integration of complex networks by hypergraph-based
  spectral clustering.
\newblock \emph{Physical Review E} \textbf{86}:056111 (2012).

\bibitem{zhu2004}
Zhu J \emph{et~al.}
\newblock An integrative genomics approach to the reconstruction of gene
  networks in segregating populations.
\newblock \emph{Cytogenet Genome Res} \textbf{105}:363--374 (2004).

\bibitem{chen2007harnessing}
Chen LS, Emmert-Streib F and Storey JD.
\newblock Harnessing naturally randomized transcription to infer regulatory
  relationships among genes.
\newblock \emph{Genome Biology} \textbf{8}:R219 (2007).

\bibitem{aten2008using}
Aten JE \emph{et~al.}
\newblock Using genetic markers to orient the edges in quantitative trait
  networks: the {NEO} software.
\newblock \emph{BMC Systems Biology} \textbf{2}:34 (2008).

\bibitem{schadt2005integrative}
Schadt EE \emph{et~al.}
\newblock An integrative genomics approach to infer causal associations between
  gene expression and disease.
\newblock \emph{Nature Genetics} \textbf{37}:710--717 (2005).

\bibitem{neto2008inferring}
Neto EC \emph{et~al.}
\newblock Inferring causal phenotype networks from segregating populations.
\newblock \emph{Genetics} \textbf{179}:1089--1100 (2008).

\bibitem{neto2013modeling}
Neto EC \emph{et~al.}
\newblock Modeling causality for pairs of phenotypes in system genetics.
\newblock \emph{Genetics} \textbf{193}:1003--1013 (2013).

\bibitem{millstein2009disentangling}
Millstein J \emph{et~al.}
\newblock Disentangling molecular relationships with a causal inference test.
\newblock \emph{BMC Genetics} \textbf{10}:23 (2009).

\bibitem{smith2003mendelian}
Smith GD and Ebrahim S.
\newblock `mendelian randomization': can genetic epidemiology contribute to
  understanding environmental determinants of disease?
\newblock \emph{International Journal of Epidemiology} \textbf{32}:1--22
  (2003).

\bibitem{li2010critical}
Li Y \emph{et~al.}
\newblock Critical reasoning on causal inference in genome-wide linkage and
  association studies.
\newblock \emph{Trends in Genetics} \textbf{26}:493--498 (2010).

\bibitem{Friedman:1999}
Friedman N, Nachman I and Pe{\'e}r D.
\newblock Learning bayesian network structure from massive datasets: The
  ``sparse candidate'' algorithm.
\newblock In \emph{Proceedings of the Fifteenth Conference on Uncertainty in
  Artificial Intelligence}, UAI'99, 206--215  (Morgan Kaufmann Publishers Inc.,
  San Francisco, CA, USA1999).

\bibitem{Friedman:2000}
Friedman N \emph{et~al.}
\newblock Using bayesian networks to analyze expression data.
\newblock \emph{Journal of Computational Biology} \textbf{7}:601--620 (2000).

\bibitem{koller2009}
Koller D and Friedman N.
\newblock \emph{Probabilistic Graphical Models: Principles and Techniques}
  ({The MIT Press}2009).

\bibitem{schmidt2007learning}
Schmidt M, Niculescu-Mizil A and Murphy K.
\newblock Learning graphical model structure using {L1}-regularization paths.
\newblock In \emph{AAAI}, vol.~7, 1278--1283 (2007).

\bibitem{zhu2008}
Zhu J \emph{et~al.}
\newblock Integrating large-scale functional genomic data to dissect the
  complexity of yeast regulatory networks.
\newblock \emph{Nature Genetics} \textbf{40}:854--861 (2008).

\bibitem{zhu2012stitching}
Zhu J \emph{et~al.}
\newblock Stitching together multiple data dimensions reveals interacting
  metabolomic and transcriptomic networks that modulate cell regulation.
\newblock \emph{PLoS Biology} \textbf{10}:e1001301 (2012).

\bibitem{zhang2013integrated}
Zhang B \emph{et~al.}
\newblock {{I}ntegrated systems approach identifies genetic nodes and networks
  in late-onset {A}lzheimer's disease}.
\newblock \emph{Cell} \textbf{153}:707--720 (2013).

\bibitem{neto2010causal}
Neto EC \emph{et~al.}
\newblock Causal graphical models in systems genetics: a unified framework for
  joint inference of causal network and genetic architecture for correlated
  phenotypes.
\newblock \emph{The Annals of Applied Statistics} \textbf{4}:320 (2010).

\bibitem{scutari2014multiple}
Scutari M \emph{et~al.}
\newblock Multiple quantitative trait analysis using {B}ayesian networks.
\newblock \emph{Genetics} \textbf{198}:129--137 (2014).

\bibitem{langfelder2007eigengene}
Langfelder P and Horvath S.
\newblock Eigengene networks for studying the relationships between
  co-expression modules.
\newblock \emph{BMC Systems Biology} \textbf{1}:54 (2007).

\bibitem{friedman1999data}
Friedman N, Goldszmidt M and Wyner A.
\newblock Data analysis with {B}ayesian networks: A bootstrap approach.
\newblock In \emph{Proceedings of the Fifteenth conference on Uncertainty in
  artificial intelligence}, 196--205  (Morgan Kaufmann Publishers Inc.1999).

\bibitem{segal2003}
Segal E \emph{et~al.}
\newblock Module networks: identifying regulatory modules and their
  condition-specific regulators from gene expression data.
\newblock \emph{Nat Genet} \textbf{34}:166--167 (2003).

\bibitem{friedman2004}
Friedman N.
\newblock Inferring cellular networks using probabilistic graphical models.
\newblock \emph{Science} \textbf{308}:799--805 (2004).

\bibitem{lee2006}
Lee S \emph{et~al.}
\newblock {{I}dentifying regulatory mechanisms using individual variation
  reveals key role for chromatin modification}.
\newblock \emph{Proc Natl Acad Sci USA} \textbf{103}:14062--14067 (2006).

\bibitem{zhang2010bayesian}
Zhang W \emph{et~al.}
\newblock A {B}ayesian partition method for detecting pleiotropic and epistatic
  {eQTL} modules.
\newblock \emph{PLoS Computational Biology} \textbf{6}:e1000642 (2010).

\bibitem{joshi2009}
Joshi A \emph{et~al.}
\newblock Module networks revisited: computational assessment and
  prioritization of model predictions.
\newblock \emph{Bioinformatics} \textbf{25}:490--496 (2009).

\bibitem{talukdar2015}
Talukdar H \emph{et~al.}
\newblock Cross-tissue regulatory gene networks in coronary artery disease.
\newblock \emph{Cell Systems} \textbf{2}:196--208  (2016).

\bibitem{ashb00}
Ashburner M \emph{et~al.}
\newblock {{G}ene ontology: tool for the unification of biology. {T}he {G}ene
  {O}ntology {C}onsortium}.
\newblock \emph{Nat Genet} \textbf{25}:25--29 (2000).

\bibitem{furey2012chip}
Furey TS.
\newblock {ChIP}--seq and beyond: new and improved methodologies to detect and
  characterize protein--{DNA} interactions.
\newblock \emph{Nature Reviews Genetics} \textbf{13}:840--852 (2012).

\bibitem{encode2012}
{The ENCODE Project Consortium}.
\newblock {{A}n integrated encyclopedia of {D}{N}{A} elements in the human
  genome}.
\newblock \emph{Nature} \textbf{489}:57--74 (2012).

\bibitem{kundaje2015integrative}
Kundaje A \emph{et~al.}
\newblock Integrative analysis of 111 reference human epigenomes.
\newblock \emph{Nature} \textbf{518}:317--330 (2015).

\bibitem{gerstein2010integrative}
Gerstein M \emph{et~al.}
\newblock Integrative analysis of the {C}aenorhabditis elegans genome by the
  {modENCODE} project.
\newblock \emph{Science} \textbf{330}:1775--1787 (2010).

\bibitem{roy2010identification}
Roy S \emph{et~al.}
\newblock Identification of functional elements and regulatory circuits by
  {D}rosophila {modENCODE}.
\newblock \emph{Science} \textbf{330}:1787--1797 (2010).

\bibitem{yue2014comparative}
Yue F \emph{et~al.}
\newblock A comparative encyclopedia of {DNA} elements in the mouse genome.
\newblock \emph{Nature} \textbf{515}:355--364 (2014).

\bibitem{chatr2014biogrid}
Chatr-aryamontri A \emph{et~al.}
\newblock The {BioGRID} interaction database: 2015 update.
\newblock \emph{Nucleic acids research} gku1204 (2014).

\bibitem{lu2007}
Lu P \emph{et~al.}
\newblock Absolute protein expression profiling estimates the relative
  contributions of transcriptional and translational regulation.
\newblock \emph{Nature Biotech} \textbf{25}:117--124 (2007).

\bibitem{schwanhausser2011}
Schwanhausser B \emph{et~al.}
\newblock {{G}lobal quantification of mammalian gene expression control}.
\newblock \emph{Nature} \textbf{473}:337--342 (2011).

\bibitem{wu2013variation}
Wu L \emph{et~al.}
\newblock Variation and genetic control of protein abundance in humans.
\newblock \emph{Nature} \textbf{499}:79--82 (2013).

\bibitem{cenik2015integrative}
Cenik C \emph{et~al.}
\newblock Integrative analysis of rna, translation and protein levels reveals
  distinct regulatory variation across humans.
\newblock \emph{Genome Research} doi:10.1101/gr.193342.115 (2015).

\bibitem{waszak2015population}
Waszak SM \emph{et~al.}
\newblock Population variation and genetic control of modular chromatin
  architecture in humans.
\newblock \emph{Cell} \textbf{162}:1039--1050 (2015).

\bibitem{grubert2015genetic}
Grubert F \emph{et~al.}
\newblock Genetic control of chromatin states in humans involves local and
  distal chromosomal interactions.
\newblock \emph{Cell} \textbf{162}:1051--1065 (2015).

\bibitem{rao20143d}
Rao SS \emph{et~al.}
\newblock A {3D} map of the human genome at kilobase resolution reveals
  principles of chromatin looping.
\newblock \emph{Cell} \textbf{159}:1665--1680 (2014).

\bibitem{cusanovich2014functional}
Cusanovich DA \emph{et~al.}
\newblock The functional consequences of variation in transcription factor
  binding.
\newblock \emph{PLoS Genetics} \textbf{10}:e1004226 (2014).

\bibitem{add1}
Qu K \emph{et~al.}
\newblock Integrative genomic analysis by interoperation of bioinformatics tools in GenomeSpace.
\newblock \emph{Nature Methods} \textbf{13}:245--247 (2016).

\end{thebibliography}
\end{document}